\documentclass[showpacs,preprintnumbers,twocolumn,superscriptaddress,floatfix]{revtex4}
\usepackage{amsmath}
\usepackage{amsfonts}
\usepackage{amssymb}
\usepackage{natbib}
\usepackage{bibentry}
\usepackage{trig}
\usepackage{keyval}
\usepackage{graphicx}
\usepackage{verbatim}
\begin{document}

\title{Strange baryon resonance production in $\sqrt{s_{NN}} = 200$ GeV $p+p$ and $Au+Au$ collisions}
\affiliation{Argonne National Laboratory, Argonne, Illinois 60439}
\affiliation{University of Birmingham, Birmingham, United Kingdom}
\affiliation{Brookhaven National Laboratory, Upton, New York 11973}
\affiliation{California Institute of Technology, Pasadena,
California 91125} \affiliation{University of California, Berkeley,
California 94720} \affiliation{University of California, Davis,
California 95616} \affiliation{University of California, Los
Angeles, California 90095} \affiliation{Carnegie Mellon University,
Pittsburgh, Pennsylvania 15213} \affiliation{University of Illinois,
Chicago} \affiliation{Creighton University, Omaha, Nebraska 68178}
\affiliation{Nuclear Physics Institute AS CR, 250 68
\v{R}e\v{z}/Prague, Czech Republic} \affiliation{Laboratory for High
Energy (JINR), Dubna, Russia} \affiliation{Particle Physics
Laboratory (JINR), Dubna, Russia} \affiliation{University of
Frankfurt, Frankfurt, Germany} \affiliation{Institute of Physics,
Bhubaneswar 751005, India} \affiliation{Indian Institute of
Technology, Mumbai, India} \affiliation{Indiana University,
Bloomington, Indiana 47408} \affiliation{Institut de Recherches
Subatomiques, Strasbourg, France} \affiliation{University of Jammu,
Jammu 180001, India} \affiliation{Kent State University, Kent, Ohio
44242} \affiliation{Institute of Modern Physics, Lanzhou, China}
\affiliation{Lawrence Berkeley National Laboratory, Berkeley,
California 94720} \affiliation{Massachusetts Institute of
Technology, Cambridge, MA 02139-4307}
\affiliation{Max-Planck-Institut f\"ur Physik, Munich, Germany}
\affiliation{Michigan State University, East Lansing, Michigan
48824} \affiliation{Moscow Engineering Physics Institute, Moscow
Russia} \affiliation{City College of New York, New York City, New
York 10031} \affiliation{NIKHEF and Utrecht University, Amsterdam,
The Netherlands} \affiliation{Ohio State University, Columbus, Ohio
43210} \affiliation{Panjab University, Chandigarh 160014, India}
\affiliation{Pennsylvania State University, University Park,
Pennsylvania 16802} \affiliation{Institute of High Energy Physics,
Protvino, Russia} \affiliation{Purdue University, West Lafayette,
Indiana 47907} \affiliation{Pusan National University, Pusan,
Republic of Korea} \affiliation{University of Rajasthan, Jaipur
302004, India} \affiliation{Rice University, Houston, Texas 77251}
\affiliation{Universidade de Sao Paulo, Sao Paulo, Brazil}
\affiliation{University of Science \& Technology of China, Hefei
230026, China} \affiliation{Shanghai Institute of Applied Physics,
Shanghai 201800, China} \affiliation{SUBATECH, Nantes, France}
\affiliation{Texas A\&M University, College Station, Texas 77843}
\affiliation{University of Texas, Austin, Texas 78712}
\affiliation{Tsinghua University, Beijing 100084, China}
\affiliation{Valparaiso University, Valparaiso, Indiana 46383}
\affiliation{Variable Energy Cyclotron Centre, Kolkata 700064,
India} \affiliation{Warsaw University of Technology, Warsaw, Poland}
\affiliation{University of Washington, Seattle, Washington 98195}
\affiliation{Wayne State University, Detroit, Michigan 48201}
\affiliation{Institute of Particle Physics, CCNU (HZNU), Wuhan
430079, China} \affiliation{Yale University, New Haven, Connecticut
06520} \affiliation{University of Zagreb, Zagreb, HR-10002, Croatia}

\author{B.I.~Abelev}\affiliation{Yale University, New Haven, Connecticut 06520}
\author{M.M.~Aggarwal}\affiliation{Panjab University, Chandigarh 160014, India}
\author{Z.~Ahammed}\affiliation{Variable Energy Cyclotron Centre, Kolkata 700064, India}
\author{J.~Amonett}\affiliation{Kent State University, Kent, Ohio 44242}
\author{B.D.~Anderson}\affiliation{Kent State University, Kent, Ohio 44242}
\author{M.~Anderson}\affiliation{University of California, Davis, California 95616}
\author{D.~Arkhipkin}\affiliation{Particle Physics Laboratory (JINR), Dubna, Russia}
\author{G.S.~Averichev}\affiliation{Laboratory for High Energy (JINR), Dubna, Russia}
\author{Y.~Bai}\affiliation{NIKHEF and Utrecht University, Amsterdam, The Netherlands}
\author{J.~Balewski}\affiliation{Indiana University, Bloomington, Indiana 47408}
% ...

\collaboration{STAR Collaboration}\noaffiliation

\date{\today}

\begin{abstract}
We report the measurements of $\Sigma (1385)$ and $\Lambda (1520)$
production in $p+p$ and $Au+Au$ collisions at $\sqrt{s_{NN}} =
200$ GeV from the STAR collaboration. The yields and the $p_{T}$
spectra are presented and discussed in terms of chemical and
thermal freeze-out conditions and compared to model predictions.
Thermal and microscopic models do not adequately describe the
yields of all the resonances produced in central $Au+Au$
collisions. Our results indicate that there may be a time-span
between chemical and thermal freeze-out during which elastic
hadronic interactions occur.
\end{abstract}

\maketitle

In ultra-relativistic heavy ion collisions, hot and dense nuclear
matter (a fireball) is created \cite{Phenix1,Star1}. When the
energy density of the created fireball is very high, deconfinement
of partons is expected to occur and a new phase of matter, the
Quark Gluon Plasma (QGP) forms. After hadronization of the QGP,
but before the interactions of the hadrons cease, the physical
properties of resonances, such as their in vacuo masses and
widths, might be modified by the density of the surrounding
nuclear medium~\cite{lut02}. In addition, the yield of resonances
might change.

The temperature and the density of the fireball  reduces as the
fireball expands. Chemical freeze-out is reached when hadrons stop
interacting inelastically. Elastic interactions continue until
thermal freeze-out. Due to their short lifetimes, a fraction of
resonances can decay before the thermal freeze-out. Elastic
interactions of the decay products with other particles in the
medium (re-scattering) may modify their momenta enough that the
parent particle can no longer be identified. The pseudo-elastic
hadronic interactions (regeneration) may increase the resonance
yields (e.g., $\Lambda + \pi \rightarrow \Sigma(1385)$)
\cite{urqmd,shuryak,rapp,kstar130}. The overall net effect of
re-scattering and regeneration on the total observed yields
depends on the time-span between chemical and thermal freeze-out,
the lifetime of the resonances and the magnitudes of the
interaction cross-sections of the decay particles
\cite{tor01,ble02}. Thermal models provide the resonance to stable
particle ratios at the chemical freeze-out. Deviations from these
predicted ratios due to re-scattering of the resonance decay
particles can be used to estimate the time-span between chemical
and thermal freeze-out.

We report on the first measurements of the production of the
$\Sigma(1385)$~\cite{sevil} and $\Lambda(1520)$~\cite{ludo} in
$p+p$ and $Au+Au$ collisions at $\sqrt{s_{NN}}=200 $ GeV. The
effects of the extended nuclear medium on the resonance yields and
momentum spectra are studied by comparing those results from the
different collision systems. Microscopic transport~\cite{urqmd}
and thermal~\cite{pbm01,bec02,starwhite} models are used to
investigate the time-span of hadronically interacting phase.

The STAR detector system \cite{bbc}, with its large time
projection chamber (TPC), is used to identify the decay products
of the $\Sigma (1385) \rightarrow \Lambda + \pi$ and $\Lambda
(1520) \rightarrow p + K$. For $Au+Au$ collisions, the number of
charged particles in the TPC is used to select the centrality of
inelastic interactions. Different $y$ and centrality selections
are necessary for $\Sigma (1385)$ and $\Lambda (1520)$ in order to
optimize the statistical significance of each measurement.

The topological reconstruction of resonance decay vertices is not
possible due to their short lifetimes resulting from their strong
decay. Instead an invariant mass calculation from the decay
daughter candidates is performed. Charged particles are identified
by the energy loss per unit length, $dE/dx$, and the momentum
measured with the TPC. The decay topology information is used to
identify the neutral $\Lambda$~\cite{multibaryon}. A large source
of background in the invariant mass spectra for both
$\Sigma(1385)$ and $\Lambda (1520)$ comes from uncorrelated pairs.
A mixed event technique, where no correlations are possible, is
used to estimate the contribution of the background~\cite{zhang}.
The background is normalized over a wide kinematic range and then
subtracted from the invariant mass distribution. For the $\Sigma
^{-}(1385)$, a $\Xi ^{-}$ peak remains as it has the same $\Lambda
+ \pi ^{-}$ decay channel. In order to enhance the statistics for
the $\Sigma ^{*}$, two charged channels are combined ($\Sigma
^{\pm }(1385)$) for $p+p$ and all four charged channels ($\Sigma
^{ \pm}(1385)$+$\overline{\Sigma} ^ {\pm }(1385)$) for $Au+Au$
collisions. Similarly for the $\Lambda ^{*}$, $\Lambda (1520)$ and
$ \overline{\Lambda} (1520)$ are combined in $p+p$ collisions. As
the $ \overline{\Lambda} (1520)$ is not observed in central
$Au+Au$ collisions, it is not included in our definition of
$\Lambda ^{*}$ in $Au+Au$.

\begin{figure}[b!]
 \centering
 $\begin{array}{cc}
\includegraphics[width=0.43\textwidth]{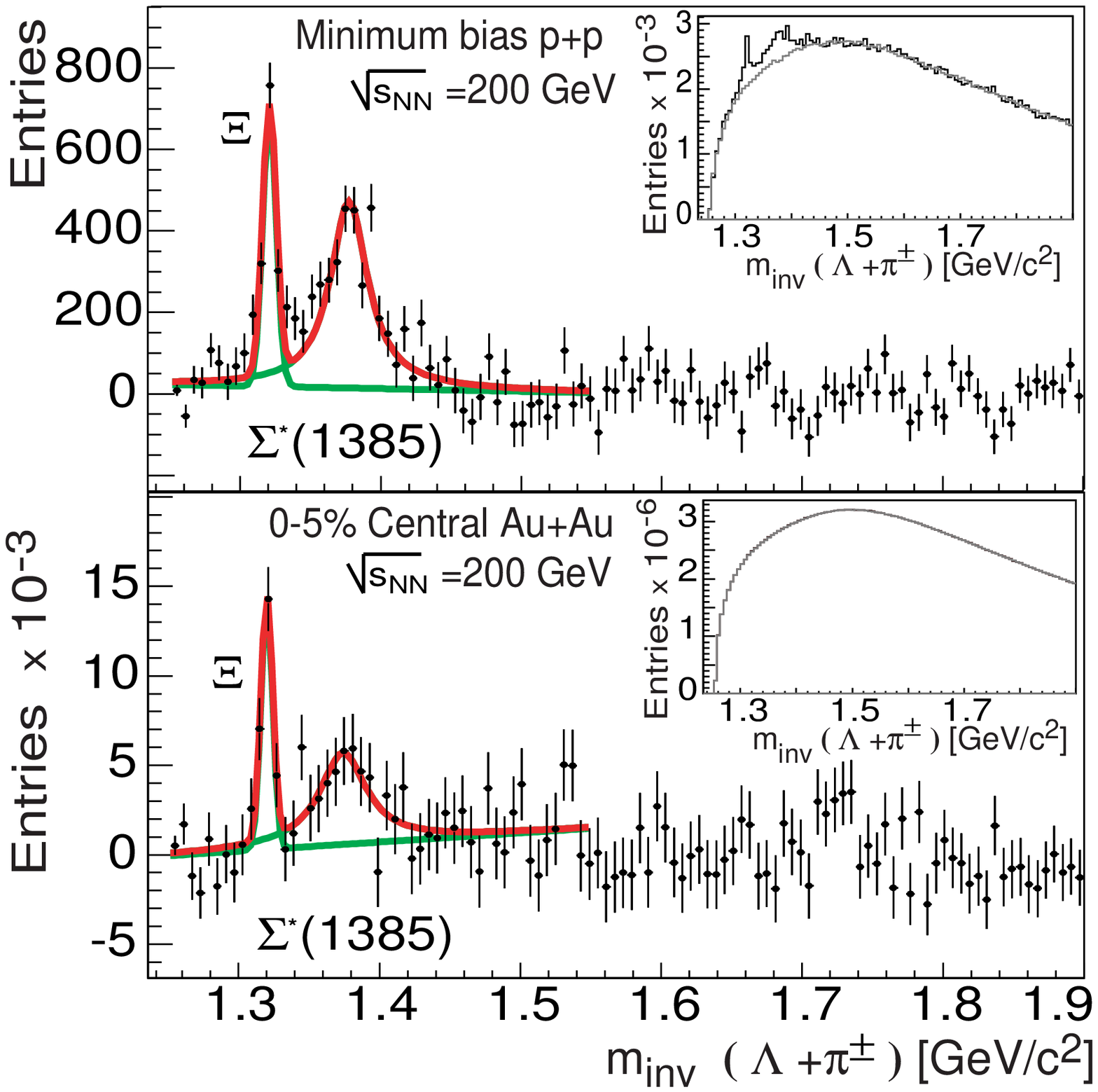}\\
\includegraphics[width=0.43\textwidth,height=7.2cm]{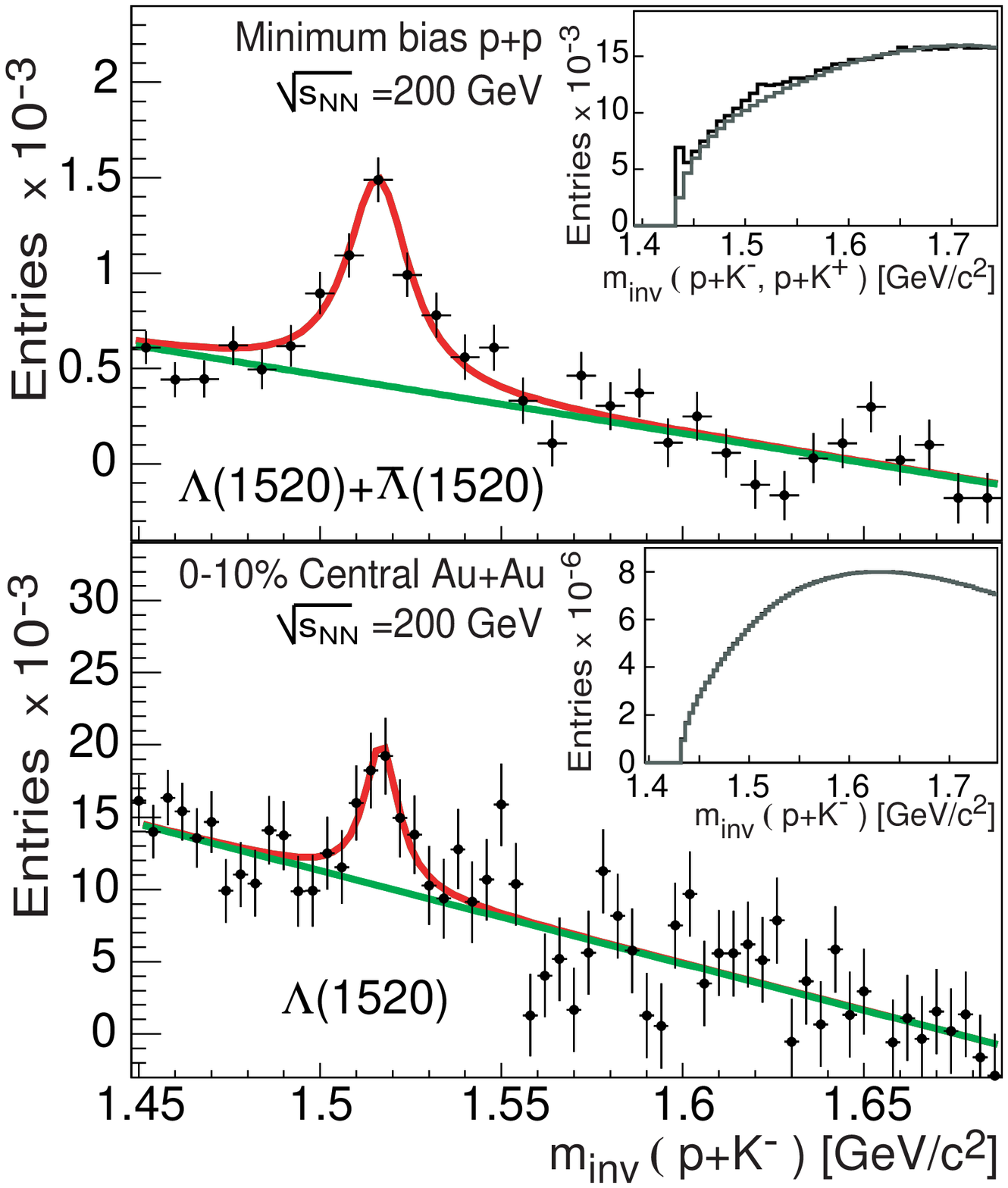}
\end{array}$
\caption{Invariant mass distributions of $\Sigma^{*}$ and
$\Lambda^{*}$ in $p+p$ and $Au+Au$ collisions at
$\sqrt{s_{NN}}=200\;\rm GeV$ before (inset) and after mixed-event
background subtraction.}
 \label{invmass}
\end{figure}

\begin{table}[b!]
\centering \caption{ Mass ($M$) and width ($\Gamma$) fit
parameters of particles from Fig.~\ref{invmass}, including
statistical and systematic errors.} \label{table1}
\begin{ruledtabular}
\begin{tabular}{lcccc} Particle & $M$ [MeV/c$^{2}$]  &
$\Gamma$ [MeV/c$^{2}$] & $p_{T}$ [GeV/c] & $|y|$ \\ \hline
 $\Xi^{-}_{(p+p)}$ & $1320\pm 1 \pm 1$   & $7 \pm 1 \pm 1$ & 0.25 -- 3.50 & $\leq 0.75$ \\
$\Xi^{-}_{(Au+Au)}$ &  $1320 \pm 1 \pm 1$ & $4 \pm 1 \pm 1$ & 0.50 -- 3.50 & $\leq 0.75$\\
$\Sigma ^{*}_{(p+p)}$ & $1376 \pm 3 \pm 3$   & $44 \pm 8 \pm 8$ & 0.25 -- 3.50 & $\leq 0.75$\\
$\Sigma ^{*}_{(Au+Au)}$ &  $1375 \pm 5 \pm 3$ & $43 \pm 5 \pm 6$ & 0.50 -- 3.50 & $ \leq 0.75$\\
 $\Lambda ^{*}_{(p+p)}$& $1516 \pm 2 \pm 2$ & $20 \pm 4 \pm 2$& 0.20 -- 2.20 & $\leq 0.50$\\
$\Lambda ^{*}_{(Au+Au)}$ & $1516 \pm 2 \pm 2$& $12 \pm 6 \pm 3$ & 0.90 -- 2.00 & $\leq 1.00$ \\
\end{tabular}
\end{ruledtabular}
\end{table}

Fig.~\ref{invmass} shows the invariant mass distributions for
$\Sigma^{*}$ and $\Lambda^{*}$ in 10 million minimum bias $p+p$
and 1.6 million central $Au+Au$ collisions. The mass ($M$) and the
width ($\Gamma$) fit parameters of the measured transverse
momentum ($p_{T}$) and rapidity ($y$) ranges are shown in
Table~\ref{table1}. These parameters and their uncertainties are
obtained from combined fits. A Gaussian distribution takes into
account the detector resolution effects on the $\Xi^{-}$. Since
the natural width dominates over the detector resolutions for both
the $\Sigma^{*}$ and $\Lambda^{*}$, a non-relativistic
Breit-Wigner distribution is used. Finally the remaining residual
background is described by a linear function. The measured widths,
taking into account the detector resolution, are, within their
uncertainties, in agreement with the PDG \cite{pdg98}. The
observed mass and the width of the $\Xi^{-}$ peak is in agreement
with the one obtained via the topological method
\cite{multibaryon}. While the masses of $\Xi$ and $\Lambda^{*}$
are also in agreement with the PDG values, there is a small
difference in the mass of the $\Sigma^{*}$. Due to limited
statistics, it is not possible to investigate this effect further.
The systematic errors include the uncertainty due to bin size
fluctuations, the normalization of the mixed event background and
the uncertainty of the straight line fit range due to correlations
in misidentified decay particles. Event and track selections were
also varied.

\begin{table}[t!]
 \centering
 \caption{$\langle p_{T}\rangle$ and yields from fits to the
$p_{T}$ spectra, $dN/dy$ for $\Lambda^{*}$ in $Au+Au$ using a
fixed $T$. The $p+p$ yields are from non-singly diffractive
collisions. $\Sigma^{*}$ represents $\Sigma^{*+} + \Sigma^{*-}$. }
 \label{table2}
\begin{ruledtabular}\begin{tabular}{lcccc}
Particle & Collision & $\langle p_{T}\rangle$ [GeV/c]   & $(dN/dy)|_{y=0}$ \\
  \hline
$\Sigma^{*}$ &  $pp_{ minbias}$ &1.02$\pm$0.02$\pm$0.07   &(10.7$\pm$0.4$\pm$1.4)$\times10^{-3}$  \\
$\overline \Sigma^{*}$ &  $pp_{ minbias}$ & 1.01$\pm$0.01$\pm$0.06    &(8.9$\pm$0.4$\pm$1.2)$\times10^{-3}$  \\
 $\overline{\Sigma}^{*}$+$\Sigma^{*}$ &  $AuAu_{ 0-5\%}$ &1.28$\pm$0.15$\pm$0.09&9.3$\pm$1.4$\pm$1.2 \\
$\overline{\Lambda}^{*}$+$\Lambda^{*}$ &   $pp_{ minbias}$ & 1.08$\pm$0.09$\pm$0.05&(6.9$\pm$0.5$\pm$1.0)$\times10^{-3}$ \\
$\Lambda ^{*}$  &  $AuAu_{0-10\%}$     & 1.20$\pm$0.20$_{\it fixed}$  &(6.3$\pm$2.1$\pm$0.8)$\times10^{-1}$  \\
$\overline{\Lambda}^{*}$+$\Lambda^{*}$ &  $AuAu_{60-80\%}$ & 1.20$\pm$0.20$_{\it fixed}$ & (8.9$\pm$2.9$\pm$1.1)$\times10^{-2}$  \\
\end{tabular}\end{ruledtabular}
\end{table}

\begin{figure}[b!]
\includegraphics[width=0.47\textwidth]{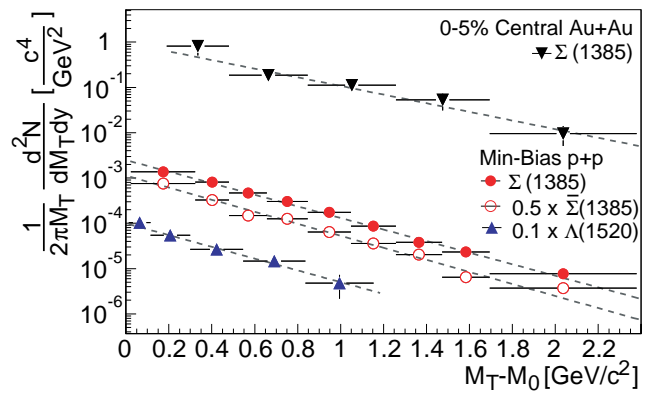}
 \caption{The transverse mass spectra for
$\Sigma^{*}$ and $\Lambda ^{*}$ in $p+p$ and in central $Au+Au$
collisions at $\sqrt{s_{NN}}= 200 \;\rm GeV$. Statistical and
systematical errors are included.}
 \label{mtspectrum}
\end{figure}

To obtain the integrated raw yields of $\Sigma ^{*}$ and
$\Lambda^{*}$, the background subtracted invariant mass spectrum
in each $p_{T}$ bin is fitted. In the corresponding mass range,
the content of each bin above the linear background fit is counted
to extract the raw yields. Monte-Carlo simulated resonances are
embedded into real $p+p$ and $Au+Au$ events to determine the
correction factors for the detector acceptance and reconstruction
efficiency. These are applied to the data and the corrected
transverse mass spectra of $\Sigma ^{*}$ and $\Lambda ^{*}$ in
$p+p$ and $Au+Au$ collisions are shown in Fig.~\ref{mtspectrum}.
The dashed curves represent an exponential fit to the
data~\cite{zhang}. The mean $p_{T}$ ($\langle p_{T}\rangle$) and
the yields at mid-rapidity ($dN/dy$) as obtained from the fit are
listed in Table~\ref{table2} together with their corresponding
statistical uncertainties. The yields are obtained by
extrapolating the fit to all $p_{T}$. The measured $p_{T}$ range
contains $85\%$ for $\Sigma ^{*}$ and $50\%$ for $\Lambda^{*}$ in
$Au+Au$ and $91\%$ for $\Sigma ^{*}$ and $\Lambda^{*}$ in $p+p$ of
the total mid-rapidity yields. For $\Lambda ^{*}$, due to the low
statistics in $Au+Au$ collisions, an inverse slope of
$T\;=400$~MeV is assumed in order to extract the particle yield.
The systematic error includes a $\Delta T\;=100$~MeV variation.
The ratio of $\overline{\Lambda}^{*}/\Lambda ^{*}=0.93\pm0.11$ in
$p+p$ collisions is extracted from the corrected yields.
Statistical limitations require that the
$\overline{\Sigma}^{*}/\Sigma ^{*} = 0.87 \pm 0.18$ in $Au+Au$
collisions are determined from the raw yields. The proximity of
these ratios to unity, reflects a small net baryon number at
mid-rapidity of both systems.

A linear increase of $\langle p_{T} \rangle$ as a function of
particle mass up to 1 $\rm GeV/c^2$ is observed in $Au+Au$ and
$p+p$ collisions \cite{multibaryon,pppaper}. The measured $\langle
p_{T}\rangle$ of $\Sigma^{*}$ and $\Lambda^{*}$ in $p+p$
collisions follow a steeper increase, similar to the trend of
heavier mass particles ($\rm > 1 \;GeV/c^{2}$). This might be due
to the fact that the higher mass particles come from events with
average multiplicities a factor of 2 or more higher than those for
the minimum bias events. The increase in the $\langle p_{T}
\rangle$ and the larger event multiplicities imply that these
resonances come from mini-jet like events \cite{dumitru}. The
re-scattering and regeneration is expected to change the $\langle
p_{T}\rangle$ in $Au+Au$ collisions. However, it is surprising
that the $\langle p_{T}\rangle$ of $\Sigma^{*}$ in $p+p$ and
$Au+Au$ collisions are in agreement within their uncertainties.

\begin{figure}[h]
 \centering
\includegraphics[width=0.43\textwidth]{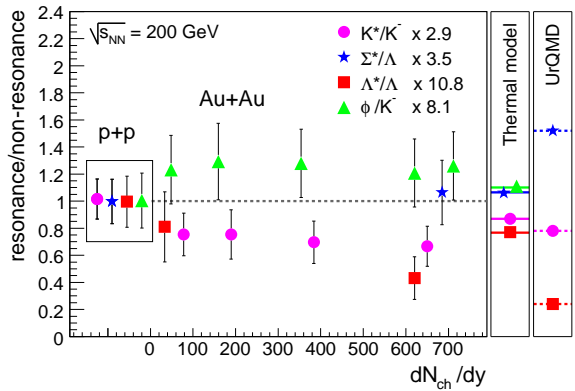}
 \caption{Resonance to stable particle ratios for
$p+p$ and $Au+Au$ collisions. The ratios are normalized to unity in
$p+p$ and compared to thermal and UrQMD model predictions for
central $Au+Au$~\cite{pbm01,ble02}. Statistical and systematic
uncertainties are included in the error bars.
 } \label{ratio}
\end{figure}

The ratios of yields of resonances to stable particles as a
function of the charged particle multiplicity are presented in
Fig.~\ref{ratio}. The ratios are normalized to unity in $p+p$
collisions to study variations in $Au+Au$ relative to $p+p$. We
measure a suppression for $\Lambda ^{*}/\Lambda$ when comparing
central $Au+Au$ with minimum bias $p+p$. $K ^{*}/K^{-}$
~\cite{zhang} seems to show a smaller suppression while the
$\Sigma ^{*} /\Lambda$, and $ \phi /K^{-}$ ~\cite{ma} ratios are
consistent with unity. In a thermal model, the measured ratios of
resonance to non-resonant particles with identical valence quarks
are particularly sensitive to the chemical freeze-out temperature,
as all of the quark content dependencies cancel out. Thermal
models require a chemical freeze-out temperature in the range
$T\;=160-180$~MeV and a baryo-chemical potential $\rm \mu _{B}=
20-50$~MeV in 200 GeV $Au+Au$ collisions to describe the stable
particle ratios~\cite{pbm01,bec02}. While these models predict the
measured $\Sigma^{*}/\Lambda$ ratio correctly within the errors,
they yield a higher ratio than the measured $\Lambda ^{*}/\Lambda$
in the most central $Au+Au$ collisions. This suggests an extended
hadronic phase of elastic and pseudo-elastic interactions after
chemical freeze-out, where re-scattering of resonance decay
particles and regeneration of resonances  will occur. The measured
resonance yields thus depend on the time-span between chemical and
kinetic freeze-out, their cross sections for re-scattering and
regeneration, and their lifetimes. The suppressed $\Lambda
^{*}/\Lambda$ and $K^{*}/K^{-}$ ratios in $Au+Au$ suggest that
re-scattering dominates regeneration in the hadronic medium after
chemical freeze-out.

A thermal model using an additional pure re-scattering phase, which
depends on the respective momenta of the resonance decay products,
after chemical freeze-out at $T = 160$ MeV, can be fit to the data
\cite{tor01,mar02}. The fit yields a hadronic lifetime of the source
of $\Delta \tau = 9^{+10}_{-5} $  fm/c from the
$\Lambda^{*}/\Lambda$ and $\Delta \tau = 2.5^{+1.5}_{-1}$  fm/c from
the $K^{*}/K$ ratio. The small difference between the time spans can
be explained by an enhanced regeneration cross section for the
$K^{*}$ in the medium. This theory is supported by the null
suppression of the $\Sigma^{*}/\Lambda$. The smaller lifetime of the
$\Sigma^{*}$ compared to the $\Lambda^{*}$ should lead to a larger
signal loss due to re-scattering, thus the lack of suppression
requires an enhanced regeneration probability of the $\Sigma^{*}$.
Based on the same argument the $K^{*}$ regeneration cross section
needs to be larger than that of the $\Lambda^{*}$ due to the
observed smaller suppression and shorter lifetime of the $K^{*}$
(i.e., defining $R$ as the ratio of regeneration to re-scattering
cross section, we find $R_{K+p} < R_{K+\pi} < R_{\Lambda+\pi}$ since
$c\tau_{K^{*}}< c\tau_{\Sigma^{*}}< c\tau_{\Lambda^{*}}$). A
microscopic model calculation (UrQMD) with a typical lifespan of
$\Delta \tau $ $=13 \pm 3 $~fm/c for the re-scattering and
regeneration phase, can describe $ K^{*}/K^{-}$ and $\Lambda
^{*}/\Lambda$ ratios approximately, but fails for the $\Sigma ^{*}
/\Lambda$~\cite{ble02}. The measured resonance yields in heavy-ion
collisions provide a tool to determine the strength of in-medium
hadronic cross sections and current microscopic transport models
such as UrQMD will have to be modified to account for such cross
sections~\cite{vogel}. The $\Delta \tau$ extracted from the
measurements can be used in comparison to the analysis of two-pion
intensity interferometry (HBT) in order to obtain an estimate for
the partonic lifetime. Identical particle HBT yields a time of 5-12
fm/c from the start of the collision to the kinetic freeze-out
(total source lifetime) \cite{nig05}. If one assumes the $\Lambda
^{*}$ to be least affected by regeneration then the extracted
$\Delta \tau$$>$4 fm/c is a lower limit on the hadronic source
lifetime, which is a sub-interval of the total source lifetime. The
remaining time would be a rough estimate on the partonic lifetime of
the source.

Although the re-scattering and regeneration scheme is discussed
predominantly, other methods to describe the data have been
proposed. For example, in a sudden freeze-out scenario, where the
time between the chemical and kinetic freeze-out is negligible,
the $\Lambda ^{*}/\Lambda$ suppression in $Au+Au$ with respect to
$p+p$ can be explained by the influence of the dense medium on the
production of $\Lambda^{*}$. Even though the valence quarks of the
$\Lambda ^{*}$ are in a $L=1^{-}$ state, it must decay through a
relative angular momentum $L=2$ process in order to conserve
isospin~\cite{kas05}. The high partial wave component of the
$\Lambda ^{*}$ in a dense medium can suppress its decay phase
space.

We have presented the first measurements of $\Sigma ^{*}$ and
$\Lambda ^{*}$ production in $p+p$ and $Au+Au$ collisions at
$\sqrt{s_{NN}}= 200$ GeV.  The large $\langle p_{T} \rangle$ of
the $\Sigma^{*}$ and $\Lambda^{*}$ measurements in $p+p$
collisions suggests that the heavy particle production receives a
significant contribution from jet-like events. The yields of
$\Sigma^{*}$, $\Lambda^{*}$, $\phi$ and $ K^{*}$ in $Au+Au$ in
comparison to $p+p$ collisions indicate the presence of
re-scattering and regeneration for a non-zero time-span between
chemical and kinetic freeze-out. A lower limit for the hadronic
source lifetime of $\Delta \tau >$ 4 fm/c is estimated based on a
thermal model including re-scattering.

We thank the RHIC Operations Group and RCF at BNL, and the NERSC
Center at LBNL for their support. This work was supported in part
by the Offices of NP and HEP within the U.S. DOE Office of
Science; the U.S. NSF; the BMBF of Germany; IN2P3, RA, RPL, and
EMN of France; EPSRC of the United Kingdom; FAPESP of Brazil; the
Russian Ministry of Science and Technology; the Ministry of
Education and the NNSFC of China; IRP and GA of the Czech
Republic, FOM of the Netherlands, DAE, DST, and CSIR of the
Government of India; Swiss NSF; the Polish State Committee for
Scientific Research; SRDA of Slovakia, and the Korea Sci. \& Eng.
Foundation.

\end{document}